\documentclass[12pt,showpacs,preprintnumbers,amsmath,amssymb]{revtex4-1}
\usepackage{graphicx}
\usepackage{color} 
\usepackage{bm}   

\begin{document}

\newcommand{\pderiv}[2]{\frac{\partial #1}{\partial #2}}
\newcommand{\deriv}[2]{\frac{d #1}{d #2}}
\newcommand{\eq}[1]{Eq.~(\ref{#1})}  
\newcommand{\infint}{\int \limits_{-\infty}^{\infty}}
\newcommand{\be}{\begin{equation}}
\newcommand{\ee}{\end{equation}}
\newcommand{\ben}{\begin{eqnarray}}
\newcommand{\een}{\end{eqnarray}}
\newcommand{\ovl}{\overline}

\title{Curl Forces and the Nonlinear Fokker-Planck Equation}

\vskip \baselineskip

\author{
  R.S. Wedemann$^{1}$,
  A.R. Plastino$^{2}$\footnote{E-mail address: arplastino@unnoba.edu.ar} and
  C. Tsallis$^{3,4}$
}
\affiliation{
$^{1}$Instituto de Matem\'atica e Estat\'\i stica, Universidade do Estado do Rio de Janeiro\\
      Rua S\~ao Francisco Xavier, 524, 20550-900, Rio de Janeiro, RJ, Brazil \\
$^{2}$CeBio y Secretar\'{\i}a de Investigaci\'on,
      Universidad Nacional Buenos Aires - Noroeste,
      UNNOBA-Conicet, Roque Saenz Pe\~na 456, Junin, Argentina  \\
$^{3}$Centro Brasileiro de Pesquisas F\'{\i}sicas and \\
      National Institute of Science and Technology for Complex Systems, \\
      Rua Xavier Sigaud 150, 22290-180, Rio de Janeiro - RJ,  Brazil \\
      $^{4}$Santa Fe Institute, 1399 Hyde Park Road, Santa Fe, New Mexico 87501,
USA}


\newpage

\begin{abstract}

  Nonlinear Fokker-Planck equations endowed with curl drift forces are investigated.
  The conditions under which these evolution equations admit stationary solutions,
  which are $q$-exponentials of an appropriate potential function, are determined. It is proved
  that when these stationary solutions exist, the nonlinear Fokker-Planck equations
  satisfy an $H$-theorem in terms of a free-energy like quantity involving the $S_q$
  entropy. A particular two dimensional model admitting analytical, time-dependent,
  $q$-Gaussian solutions is discussed in detail. This model describes a system of
  particles with short-range interactions, performing overdamped motion under
  drag effects due to a rotating resisting medium. It is related to models
  that have been recently applied to the study of type-II superconductors.
  The relevance of the present developments to the study of complex systems
  in physics, astronomy, and biology, is discussed.

\vskip 0.2cm



\end{abstract}
\maketitle


\section{Introduction}

 The nonlinear Fokker-Planck equation \cite{F05} constitutes a powerful tool for the
study of diverse phenomena in complex systems
\cite{RCN2015,GBT2014,MMPL2002,LAB2001,S2001,F2002,ASMNC2010}, with applications
including (among many others) type-II superconductors \cite{RNC2012},
granular media \cite{CRSA2015}, and self-gravitating systems \cite{S2003,C2003}.
It governs the behavior of a time-dependent density $F({\bm x}, t)$, where ${\bm x} \in \Re^N$
designates a location in an $N$-dimensional configuration space. The evolution
of $F$ is determined by two terms: a nonlinear diffusion \cite{FF05,U2009} term
and a linear drift term (more general equations with nonlinear drift terms
have also been proposed, but we are not going to consider them in the present work).
In several of the above mentioned applications, the density $F$ is a real physical density
(as opposed to a statistical ensemble probability density) describing the evolving
distribution of a set of interacting particles executing overdamped motion, in the
relevant configuration space \cite{ASMNC2010,CSNA2014}.
In these kind of scenarios, the nonlinear diffusion term constitutes
an effective description of the interaction between the particles, while the drift term describes
the effects of other external forces acting upon them. The nonlinear Fokker-Planck equations
recently addressed in the literature exhibit several interesting and physically relevant properties.
They obey an $H$-theorem in terms of a free-energy-like quantity \cite{SNC2007}. In some important cases,
the nonlinear Fokker-Planck equations admit exact analytical solutions of the $q$-Gaussian form,
that can be interpreted as maximum entropy ($q$-maxent) densities obtainable from the optimization under
appropriate constraints, of the nonadditive power-law entropic functionals, $S_q$ \cite{T88,T2009}.
Indeed, there is a deep connection between the nonlinear Fokker-Planck dynamics and the generalized
thermostatistics based on the $S_q$ entropies. Although this connection was first pointed out
more than twenty years ago \cite{PP95}, its full physical implications are being systematically explored
only in recent years (see, for instance, \cite{RCN2015,CSNA2014,CM2015,A2015,RN2016}
and references therein). A remarkable example of this trend is given by
experimental work on granular media published in 2015 \cite{CRSA2015}
that verifies, within a $2\%$ error and for a wide experimental range,
a scale relation predicted in 1996 on the basis of the theoretical analysis of $q$-Gaussian solutions of
the nonlinear Fokker-Planck equation \cite{TB96}. The particular case of the nonlinear Fokker-Planck
equation with vanishing drift corresponds to the porous media equation.

Virtually all the literature on the nonlinear Fokker-Planck equation
and its applications deals with Fokker-Planck equations
in which the drift forces ${\bm K}$ can be derived from a potential function $V({\bm x})$, leading to stationary
densities which are $q$-exponentials of the potential $V$. In the present contribution, we consider more general
scenarios where the drift force ${\bm K}$ has, besides a component given by minus the gradient of a potential $V$,
a term ${\tilde {\bm K}}$ that does not come from a potential. In two or three space dimensions,
this situation corresponds to having forces exhibiting a non vanishing rotational or {\it curl},
which are usually referred to as {\it curl forces} \cite{BS2012}. The incorporation of curl forces
enriches the dynamical features of the nonlinear Fokker-Planck equations, enabling it to describe a
wider set of phenomena. Curl forces, although not dynamically fundamental \cite{BS2016}, are nevertheless relevant
as useful effective descriptions of diverse physical problems, as for example the nonconservative
force fields generated by optical tweezers \cite{HTJF2009}. Dynamical systems with curl forces have
interesting properties that are not yet fully understood and are the subject of current
research~\cite{BS2016,BS2015}. In the present work, we investigate the behavior of nonlinear Fokker-Planck
equations under the presence of curl forces. We determine the conditions under which these evolution
equations admit stationary solutions of the $q$-maxent form and satisfy an $H$-theorem. We also discuss
in detail a two dimensional example admitting analytical time-dependent solutions, that describes a set
of interacting particles undergoing overdamped motion, under the drag effect arising from a uniformly rotating medium.

\section{The Nonlinear Fokker-Planck Equation}\label{sec:NonLinearFokkerPlanck}

In the present work, we shall consider nonlinear Fokker-Planck equations (NLFP)
of the form
\be
  \label{eq:NLFPE}
  \frac{\partial F}{\partial t} = D \nabla^2 [F^{2 - q}] - {\bm \nabla}\cdot [F {\bm K}]\, ,
\ee
where $F({\bm x},t)$ is a time-dependent density, $D$ is a diffusion
constant, ${\bm K}(\bm x)$ is a drift force, and $q$ is a real
parameter characterizing the (power-law) nonlinearity appearing in the diffusion term.
The density $F$ is a dimensionless quantity
of the form $F= \rho({\bm x}, t)/\rho_0$, where $\rho$ has dimensions of
inverse volume and $\rho_0$ is a constant with the same
dimensions as $\rho$. Therefore, the dimensional density
$\rho({\bm x}, t)$ obeys the evolution equation
$\partial [\rho/\rho_0]/\partial t = D \nabla^2 [(\rho/\rho_0)^{2-q}]
- {\bm \nabla} \cdot [(\rho/\rho_0) {\bm K}]$.
As already mentioned, in the most frequently studied case of Eq.(\ref{eq:NLFPE}),
the drift force ${\bm K}$ is assumed to arise from a potential function $V({\bm x})$,
\be
  \label{eq:morroabaixo}
  \bm K = - {\bm \nabla} V\, .
\ee
The stationary solutions of the NLFP then satisfy
\be
  \label{eq:StatNLFPE}
  {\bm \nabla} \Bigl[ D {\bm \nabla} \left(F^{2 - q} \right) +  F ({\bm \nabla} V) \Bigr] = 0\, .
\ee
Let us consider the $q$-statistical ansatz~\cite{T2009}
\ben
  \label{eq:TsallisF}
  F_q &=& A \exp_q[-\beta V({\bm x})] \cr
  &=& A [1 - (1-q)\beta V({\bm x})]_{+}^\frac{1}{1-q}\, ,
\een
where $A$ and $\beta$ are constants to be determined, and the function
$\exp_q(z) =[1 + (1-q)z]_{+}^\frac{1}{1-q}$, usually referred to as the $q$-exponential
function, vanishes whenever $1 +(1-q) z\le 0$.
One finds that the ansatz given by Eq.(\ref{eq:TsallisF})
complies with the equation
\be
  \label{auxistation}
  D {\bm \nabla} \left(F^{2 - q} \right) +  F ({\bm \nabla} V) = 0\, ,
\ee
if
\be
  \label{eq:AbetaRelation}
  (2-q) \beta D = A^{q-1}\, .
\ee
It therefore satisfies also equation (\ref{eq:StatNLFPE})
and constitutes a stationary solution of the NLFP equation.
In summary, the $q$-exponential ansatz (\ref{eq:TsallisF}) is
a stationary solution of the NLFP equation, if the drift force
$\bm K $ is derived from a potential and $A$ and $\beta$ satisfy
the relation (\ref{eq:AbetaRelation}).
We shall assume that the stationary distribution
$F_q$ has a finite norm, that is, $\int F_q \, d^N{\bm x} = I < \infty$.
The specific conditions required for $F_q$ to have a finite norm
(such as the allowed range of $q$-values) cannot be stated in general,
because they depend on the particular form of the potential function
$V({\bm x})$. Since in many applications the solution of the NLFP
equation is interpreted as a physical density (as opposed to a probability
density) we assume finite norm, but not necessarily normalization to unity.
The stationary density $ F_{q} $ can be regarded as a $q$-maxent distribution,
because it maximizes the nonextensive $q$-entropic functional $S_{q}$ under
the constraints corresponding to the norm and the mean value of the potential $V$
\cite{PP95,T2009}.

In the limit $q \to 1$, the standard linear Fokker-Planck equation,
\be
  \label{eq:FPE}
  \frac{\partial F}{\partial t} = D \nabla^2 F - {\bm \nabla} \cdot [ F \bm K ] \, ,
\ee
is recovered. In this limit, the $q$-maxent stationary
density (\ref{eq:TsallisF}) reduces to the exponential,
Boltzmann-Gibbs-like density,
\be
\label{eq:ProbDist_BG}
F_{BG} = \frac{1}{Z} \exp [-\frac{1}{D} V(\bm x)]\, ,
\ee
with the condition (\ref{eq:AbetaRelation}) becoming $\beta D =1$,
independent of the normalization constant $A$. The density $F_{BG}$ is normalized to one
provided that $Z= \int \exp [-\frac{1}{D} V(\bm x)] d{\bm x}$.
The density $ F_{BG}$ optimizes the Boltzmann-Gibbs entropy
$S_{BG} = -\int F \ln F d{\bm x}$, under the constraints of
normalization and the mean value $\langle V \rangle$
of the potential $V$.

Note that a dynamical system with a phase space flux of the form
(\ref{eq:morroabaixo}) (that is, of a gradient form)
evolves always {\it down-hill} on the potential energy landscape,
so as to minimize the potential energy function $V({\bm x})$.
The components $\{K_i, \,\,\, i=1, \ldots, N\}$ of such a field
satisfy
\be
\frac{\partial K_i}{\partial x_j} =  \frac{\partial K_j}{\partial x_i} =
\frac{\partial^2 V}{\partial x_i\partial x_j}\, ,
\ee
which in two or three dimensions leads to
$ \bm K \neq - {\bm \nabla} V \iff {\bm \nabla} \times {\bm K} \neq 0 $\, .

\section{Nonlinear Fokker-Planck Equation with Curl Drift Forces:
Stationary Solutions}\label{subsec:Non_Grad}

Now we consider  NLFP equations endowed with drift forces having two terms, one
exhibiting the gradient form and the other one not arising from the gradient of
a potential. That is, we consider drift fields of the form
\be
   \label{eq:K_non_gradient}
   {\bm K} = {\bm G} + {\bm {\tilde K}}\, ,
\ee
where the force $\bm G$ is equal to minus the gradient of some potential function
$V({\bm x})$, while the component ${\bm \tilde K}$ does not come from a potential
(that is, $\partial \tilde K_i /\partial x_j \ne \partial \tilde K_j /\partial x_i$).
Our aim is to determine under which conditions a density proportional to the
$q$-exponential of the potential $V$ still provides a stationary solution of the
NLFP equation, preserving thus the link between this equation and the generalized nonextensive
thermostatistics. Substituting the above dirft force $\bm K$ and $q$-exponential
density $F_q$ (\ref{eq:TsallisF}) into the stationary NLFP equation (\ref{eq:StatNLFPE}),
one obtains
\be
   \label{eq:StatNLFPE_NonGrad}
   D \nabla^2 [F_q^{2 - q}] + {\bm \nabla} \cdot [ F_q ({\bm \nabla} V) ] - {\bm \nabla} [F_q {\bm {\tilde K}}] = 0\, .
\ee
It can be verified that, if $A$ and $\beta$ satisfy (\ref{eq:AbetaRelation}), the sum of the first two terms
in the above equation vanish, since $F_q$ is a stationary solution
of the NLFP equation (\ref{eq:StatNLFPE}), when the drift field ${\bm K}$ consists solely of the
gradient field $\bm G$. In order for $F_q$ to comply also with the full NLFP equation
(\ref{eq:StatNLFPE_NonGrad}), including the drift contribution associated with the non-gradient
  field ${\tilde {\bm K}}$, it is then necessary that
\be
  \label{eq:K1_Condition}
  {\bm \nabla} [F_q {\bm {\tilde K}}] = 0\, .
\ee
If the above relation is satisfied, the density $F_q$ constitutes  a stationary solution
of the full NLFP equation, corresponding to the complete drift force
${\bm K } = -({\bm \nabla} V) + {\tilde {\bm K}}$. To have the $q$-maxent stationary solution,
one therefore requires
\be
  \label{consis}
  {\bm \nabla} \left( {\bm {\tilde K}} A[1 - (1-q)\beta V]^\frac{1}{1-q} \right) = 0 \, ,
\ee
which in turn leads to the following relation between the non-gradient drift component
${\bm {\tilde K}}$ and the potential function $V(\bm x)$
\be
  \label{eq:K1_Condition2}
  [1 - (1-q)\beta V]({\bm \nabla} \cdot {\bm {\tilde K}}) - \beta ({\bm {\tilde K}} \cdot {\bm \nabla} V) = 0 \, .
\ee
This is a consistency relation that the potential function $V$, the non-gradient
force field ${\bm {\tilde K}}$, the Lagrange multiplier $\beta$, and the entropic parameter $q$
have to satisfy, in order that the nonlinear Fokker-Planck equation admits
the $q$-maxent stationary solution (\ref{eq:TsallisF}).
The general $\beta$-dependent equation (\ref{eq:K1_Condition2}) constitutes a rather
complicated relation between the non-gradient field ${\bm {\tilde K}}$
and the potential function $V$, which is difficult to characterize. Moreover, this relation depends
explicitly on the value of $\beta$. This means that for given forms of ${\tilde {\bm K}}({\bm x})$
and $V({\bm x}$), one may have stationary solutions of the $q$-maxent form (\ref{eq:TsallisF}),
only for particular values of $\beta$.

It follows from the relation (\ref{eq:K1_Condition2}) that, in order for the NLFP equation
to admit the $\beta$-parameterized family of stationary solutions (\ref{eq:TsallisF}),
with a continuous allowed range of $\beta$-values, two conditions
have to be fulfilled. On the one hand, the non-gradient
component of the drift, ${\bm {\tilde K}}$, has to be a
divergenceless vector field,
\be
  \label{eq:K1_divergenceless}
  {\bm \nabla} \cdot {\bm {\tilde K}} = 0\, .
\ee
On the other hand, ${\bm {\tilde K}}$ has to be everywhere orthogonal
to the gradient of the potential,
\be
  \label{eq:K1_Orthogonal}
  {\bm {\tilde K}} \cdot ({\bm \nabla} V) = 0 \, .
\ee
Notice that conditions (\ref{eq:K1_divergenceless})
and (\ref{eq:K1_Orthogonal}) are not only sufficient,
but also necessary conditions for the ansatz (\ref{eq:TsallisF})
to be a stationary solution of the NLFP equation (\ref{eq:NLFPE}),
for a continuous range of $\beta$-values. Indeed, if  (\ref{eq:TsallisF})
solves  (\ref{eq:NLFPE}) for such a set of $\beta$-values,
the left hand side of (\ref{eq:K1_Condition2}), which is an inhomogeneous linear function
of $\beta$, has to vanish for an interval of values of $\beta$.
This clearly implies that both the independent term, and the coefficient of the
$\beta$-linear term, have to vanish individually, leading in turn to
conditions (\ref{eq:K1_divergenceless}) and (\ref{eq:K1_Orthogonal}).
It is intersting that these conditions do not explicitly depend
on the value of the $q$-parameter, constituting a $q$-invariant structure.
The stationary solution guaranteed by these conditions is a physical solution
when it is normalizable (otherwise, it is not physical, although still
formally a solution of the NLFP equation). The normalizability of the stationary solution
depends on the particular shape of the potential $V$ and on the value of $q$
and, as already mentioned, can only be studied in a case by case way.

In two or three space dimensions, the decomposition ${\bm K} = {\bm G} + {\tilde {\bm K}}$,
with ${\bm G} = - {\bm \nabla} V$ and ${\bm \nabla} \cdot \tilde {\bm K} = 0$, resembles the
decomposition of a vector field into a curlless (irrotational) component
and a solenoidal (divergenceless) component arising from the celebrated
Helmholtz theorem \cite{MF53}. We are not, however, imposing the boundary conditions
on the fields ${\bm K}$, ${\bm G}$, and ${\tilde {\bm K}}$, that are
usually considered in connection with the Hemlholtz decomposition. Furthermore,
we require the point to point orthogonality of the irrotational and the
divergenceless components of ${\bm K}$, which is not a condition usually considered in
connection with the Helmholtz decomposition.

It is interesting that the Helmholtz-like decomposition (\ref{eq:K_non_gradient}), with orthogonal
irrotational and divergenceless parts, ${\bm G} \cdot {\tilde {\bm K}} = 0$,
arises naturally in some circumstances. For instance, the most general
rotationally invariant vector field in two dimensions has precisely this form.
Indeed, such vector fields are of the form
\ben \label{generalrotat}
{\bm G} &=& -g(r) {\bm e}_r\, , \cr
{\tilde {\bm K}} &=& l(r) {\bm e}_{\theta}\, ,
\een
where $g(r)$ and $l(r)$ are functions of the radial
coordinate $r = (x^2 + y^2)^{1/2}$ and
${\bm e}_r$ and ${\bm e}_{\theta}$
respectively denote the radial and tangential
unit vectors. It is clear that the field ${\bm G}$ in (\ref{generalrotat}) is of
the form $-{\bm \nabla} V(r)$ with $V(r) = \int^{r^{\prime}} g(r^{\prime}) dr^{\prime}$, and that
the field ${\tilde {\bm K}}$ satisfies ${\bm \nabla} \cdot {\tilde {\bm K} = 0}$ and
${\bm G} \cdot {\tilde {\bm K}} = 0$.

Summing up, we have thus determined that the NLFP equation (\ref{eq:NLFPE}) having a non-potential drift force
of the form (\ref{eq:K_non_gradient}) admits, for a continuous range of values of the parameter
$\beta$, the family of $q$-maxent stationary solutions (\ref{eq:TsallisF}) if and only if the relations
(\ref{eq:K1_divergenceless}) and (\ref{eq:K1_Orthogonal}) are satisfied.

\section{$H$-Theorem}

We are now going to explore the possibility of formulating an $H$-theorem
for the nonlinear Fokker-Planck equations, endowed with a drift term
involving a non-vanishing-curl force ${\bm {\tilde K}}$, not derivable from the
potential function $V$. Let us first consider the time derivative of the
power law entropic functional $S_{q^*}$, with $q^* = 2-q$. This is a reasonable choice,
because $q^*$ is precisely the exponent that appears inside the Laplacian term
in the NLFP equation (\ref{eq:NLFPE}). The duality $q \rightarrow 2-q$ appears frequently in the
$q$-generalized thermostatistical formalism \cite{T2009}. We have,
\ben \label{stder}
  \frac{dS_{q^*}}{dt} &=& \frac{q^*}{1-q^*} \int F^{q^*-1} \frac{\partial F}{\partial t} d^N{\bm x}
        = D q^{*2} \int F^{2q^*-3} |{\bm \nabla} F|^2 d^N{\bm x} \cr
        &+& q \int F^{q^*-1} \left( {\bm \nabla} F \right) \cdot  \left( {\bm \nabla} V  \right) d^N{\bm x}
        + \int F^{q^*} \left({\bm \nabla} \cdot {\tilde {\bm K}} \right) d^N{\bm x}\, .
\een
It is clear that the first term in the above expression is definite positive. However,
the second term does not have a definite sign. Consequently, the time derivative of
 $S_{q^*}$ does not have a definite sign and the entropic form $S_{q^*}$ does not
 itself verify an $H$-theorem. The last two terms in the expression for
 $\frac{dS_{q^*}}{dt} $, describing the contribution of the drift term to the change in the entropy,
 suggest that a linear combination of $S_{q^*}$ and of the mean value of the potential function $V$
 may comply with an $H$-theorem. The time derivative of $\langle V \rangle = \int F V d^N{\bm x}$ is
\ben \label{vtder}
  \frac{d\langle V \rangle}{dt} &=& \int V \frac{\partial F}{\partial t} d^N{\bm x} \cr
       &=& -q^* D \int F^{q^*-1} \left( {\bm \nabla} F \right) \cdot  \left( {\bm \nabla} V  \right) d^N{\bm x} -
      \int F |{\bm \nabla} V|^2 d^N{\bm x} + \int F  \left( {\bm \nabla} V\right)\cdot {\tilde {\bm K}} d^N{\bm x}\, .
\een
Combining now equations (\ref{stder}) and (\ref{vtder}) one obtains, after some algebra,
\ben \label{svtder}
  \frac{d}{dt} \left(D S_{q^*} - \langle V \rangle  \right) &=&
  \int F \Bigl| q^* D F^{q^*-2} \left({\bm \nabla} F \right) + {\bm \nabla} V \Bigr|^2 d^N{\bm x} \cr &+&
  \int F^{q^*} \left( {\bm \nabla} \cdot {\tilde {\bm K}} \right) d^N{\bm x} +
  \int F  \left( {\bm \nabla} V\right)\cdot {\tilde {\bm K}} d^N{\bm x}\, .
\een
If the curl component ${\tilde {\bm K}}$
of the drift force complies with the requirements given by equations
(\ref{eq:K1_divergenceless}) and (\ref{eq:K1_Orthogonal}),
which are necessary and sufficient for the nonlinear Fokker-Planck equation to have the family of
$q$-maxent stationary solutions (\ref{eq:TsallisF}), it follows from (\ref{svtder})
that the nonlinear Fokker-Planck equations
satisfies the $H$-theorem,
\ben \label{hteo}
 \frac{d}{dt} \left(D S_{q^*} - \langle V \rangle  \right)
    &=&  \int F \Bigl| q^* D F^{q^*-2} \left({\bm \nabla} F \right) + {\bm \nabla} V \Bigr|^2 d^N{\bm x} \cr
    &=& \Big\langle \Bigl| q^* D F^{q^*-2} \left({\bm \nabla} F \right) + {\bm \nabla} V \Bigr|^2  \Big\rangle
   \ge 0\, .
\een
It is worth stressing that the conditions (\ref{eq:K1_divergenceless}) and (\ref{eq:K1_Orthogonal})
for having stationary $q$-maxent solutions are essentially the same as those for having an
$H$-theorem.

There is an interesting consequence of the $H$ theorem, in relation with the uniqueness
of the decomposition (\ref{eq:K_non_gradient}) of the total drift force ${\bm K}$  into a gradient
component ${\bm G} = -{\bm \nabla} V$ and an (orthogonal) divergenceless component ${\tilde {\bm K}}$.
Let us assume that that total drift force can be decomposed in this fashion in two different ways,
${\bm K} = -{\bm \nabla} V_1 + {\tilde {\bm K}}_1 = -{\bm \nabla} V_2 + {\tilde {\bm K}}_2$.
If the nonlinear Fokker-Planck
equation admits a stationary solution (of finite norm) $F_{\rm st}$, it follows
from the $H$-theorem (\ref{hteo}) that
\be
  {\bm \nabla} V_1 = {\bm \nabla} V_2 = -q^* D F_{\rm st}^{q^*-2} \left({\bm \nabla} F_{\rm st} \right)\, ,
\ee
which, in turn, implies also that ${\tilde {\bm K}}_1 = {\tilde {\bm K}}_2$.
Consequently, if the nonlinear Fokker-Planck equation admits a stationary solution,
the decomposition of the total drift force into the sum of a gradient term and a
divergenceless term is unique.

\section{Quadratic Potential and Linear Drift}

We now consider in detail the case of a quadratic potential $V$ and
a linear drift ${\bm {\tilde K}}$. We shall se that in this case
the conditions (\ref{eq:K1_divergenceless}) and (\ref{eq:K1_Orthogonal})
are required even for having a stationary solution of the $q$-exponential
form (\ref{eq:TsallisF}) for one, single value of the parameter $\beta$.
We assume and potential and a drift field respectively of the forms,
\be
  \label{eq:quad_potential}
  V(\bm x) = \sum\limits_{ij} (a_{ij} x_i x_j) + \sum\limits_i (b_i x_i)\, ,
\ee
\be
  \label{eq:linear_field}
  {\bm {\tilde K}}(\bm x) = \sum\limits_j (c_{ij} x_j) + d_i\, ,
\ee
with the $a_{ij}$, $c_{ij}$, $b_i$ and $d_i$ constant coefficients.
We can assume $a_{ij} = a_{ji}$, although the $c_{ij}$ are not necessarily symmetric.
Equation (\ref{eq:K1_Condition2}) leads to a set of constraints on these coefficients, thus defining
$V(\bm x)$ and ${\bm {\tilde K}}(\bm x)$.
If we substitute equations (\ref{eq:quad_potential}) and (\ref{eq:linear_field}) in
Eq.(\ref{eq:K1_Condition2}), we obtain
\begin{align}
  \label{eq:K1_Condition_Vquad}
     & \left\{1 - (1-q)\beta \left[ \sum\limits_{ij} (a_{ij} x_i x_j) + \sum\limits_i (b_i x_i) \right] \right\} \left( \sum\limits_{k} c_{kk} \right) \nonumber \\
     & - \beta \sum\limits_{k} \left[ \sum\limits_i (c_{ki} x_i) + d_k \right] \left[ \sum\limits_j ((a_{kj} + a_{jk}) x_j) + b_k \right] = 0 \, .
\end{align}
Equation (\ref{eq:K1_Condition_Vquad}) is a second degree polynomial in the $x_i$'s that
is equal to zero. Since this equality should hold for any value of ${\bm x}$, the coefficients
of the different powers of the $x_i$ should each be equal to zero. Therefore, by separately
equating to zero the independent zero-th, first and second order terms in the
left-hand side of Eq.(\ref{eq:K1_Condition_Vquad}), one obtains
\begin{subequations}
  \be  \sum\limits_k ( c_{kk} - \beta d_k b_k) = 0\, , \label{eq:ckkCond} \ee
  \be  \sum\limits_k \left[ (1-q) c_{kk} b_i + c_{ki} b_k + (a_{ki} + a_{ik}) d_k \right] = 0\, , \quad \forall i\, , \label{eq:FirstOrderCond} \ee
  \be
   \sum\limits_k \left[ (1-q) c_{kk} (a_{ij} + a_{ji}) + c_{ki} (a_{kj} + a_{jk}) + c_{kj} (a_{ki} + a_{ik}) \right] = 0\, , \quad \forall {i,j}\, .
   \label{eq:SecOrderCond}
  \ee
\end{subequations}
With symmetric $a_{ij}$, we shall now assume
\be
   \label{eq:zero_determinant}
   \text{det } | a_{ij} | \ne 0\, .
\ee
This assumption is also necessary if $V(\bm x)$ should represent a confining potential,
leading to a normalizable stationary state of the nonlinear Fokker-Planck equation.

If we introduce an appropriate shift in the $x_i$ coordinates, it is possible to
work using a potential $V(\bm x)$ (Eq.(\ref{eq:quad_potential})) with no linear terms.
We thus define
\be
  \ovl{x}_i = x_i - r_i \, ,
\ee
so that the $r_i$ are constants that can be derived from constraints, as we will show.
We can then express Eq.(\ref{eq:quad_potential}) in terms of the $\ovl{x}_i$ as
\be
  \label{eq:quad_potential_new}
  V(\bm x) = \sum\limits_{ij} a_{ij} [\ovl x_i \ovl x_j + (\ovl x_i r_j + \ovl x_j r_i) +
             r_i r_j ] + \sum\limits_i b_i (\ovl x_i + r_i) \, .
\ee
The linear terms in Eq.(\ref{eq:quad_potential_new}) are now
\be
  \sum\limits_i \left\{ \left[ \sum\limits_j (a_{ij} r_j + a_{ji} r_j) \right] + b_i \right\} \ovl x_i  \, ,
\ee
and they will vanish if the $r_j$'s satisfy,
\[
   b_i + \sum\limits_j (a_{ij} + a_{ji}) r_j = 0\, , \quad \text{   or}
\]
\be
  \label{eq:no_linear_terms}
  b_i + 2 \sum\limits_j a_{ij} r_j = 0 \, , \quad i = 1, \ldots , N \, .
\ee
The $N$ equations (\ref{eq:no_linear_terms}) can be solved for the $r_j$'s
because the condition in Eq.(\ref{eq:zero_determinant}) holds.
The constant term  $\left( \sum_i b_i r_i \right) + \left(\sum_{ij}a_{ij}r_ir_j \right)$
in the potential $V$ can be ignored and eliminated: since the potential enters the
NLFP equation only through its gradient, this constant term has no physical significance.
Therefore, in terms of the shifted coordinates $\ovl x_i$, we have
\begin{subequations}
\begin{align}
        V({\bm {\ovl x}}) &= \sum\limits_{ij} a_{ij} {\ovl x}_i {\ovl x}_j \, , \label{eq:quad_potential_pure}\\
        {\tilde K}_i({\bm {\ovl x}}) &= \sum\limits_j ( c_{ij} {\ovl x}_j ) + {\ovl d}_i\, , \label{eq:field_shifted_x}
\end{align}
\end{subequations}
where $\ovl d_i = \sum\limits_j (c_{ij} r_j) + d_i$. We thus see that, after an
appropriate shift in the phase space variables, the problem reduces to that of a
homogeneous, quadratic potential.

If the associated nonlinear Fokker-Planck equation admits a $q$-maxent
stationary solution, even for one single value of $\beta$,
it follows from Eq.(\ref{eq:ckkCond}) that we must have
\be
  \sum\limits_j c_{jj} = 0 \implies  {\bm \nabla} \cdot {\bm {\tilde K}} = 0,
\ee
from which it follows that the condition ${\tilde {\bf K}} \cdot {\bf \nabla}V =0$
also follows. In other words, for a quadratic potential $V$ and a linear drift
${\bm {\tilde K}}$, if one has a $q$-maxent stationary solution even for one single
value of $\beta$, it is possible after a coordinates shift to recast the
system in terms of a drift field complying with conditions
(\ref{eq:K1_divergenceless}) and (\ref{eq:K1_Orthogonal}).

\section{Two Dimensional System with Exact Time-Dependent $q$-Gaussian Solutions.}

We now consider, as an example of a time-dependent solution of a nonlinear Fokker-Planck
equation with a ${\bm {\tilde K}}$ not arising from a potential, that admits a
$q$-maxent stationary solution, a bi-dimensional system submitted to the following
quadratic potential and nongradient linear drift term. For simplicity of notation, we will name the
phase space state variables as $x \equiv {\ovl x}_1$ and $y \equiv {\ovl x}_2$, so that
the potential and drift term can be expressed as
\begin{align}
        V({\bm {\ovl x}}) &=  a (x^2 + y^2) \, , \label{eq:ex_quad_potential}\\
        {\bm {\tilde K}}({\bm {\ovl x}}) &= (-by, +bx)\, . \label{eq:ex_field_nongrad}
\end{align}
It can be verified that (\ref{eq:ex_quad_potential}) and (\ref{eq:ex_field_nongrad})
satisfy conditions given by equations (\ref{eq:K1_divergenceless}) and (\ref{eq:K1_Orthogonal}).
The NLFP equation then has the form
\be
  \label{eq:NLFPE_QuadPot}
  \frac{\partial F}{\partial t} = D \nabla^2 [F^{2 - q}] + \frac{\partial [(2ax + by)F]}{\partial x}
       + \frac{\partial [(2ay - bx)F]}{\partial y}\, .
\ee
We propose the ansatz
\be
   \label{eq:TsallisAnsatz}
   F(x, y, t) = \eta (t) \left[ 1 - (1-q)(\alpha (t) x^2 + \delta (t) xy + \gamma (t) y^2) \right]^\frac{1}{(1-q)}\, ,
\ee
where $\eta (t)$, $\alpha (t)$, $\delta (t)$ and $\gamma (t)$ are time-dependent parameters.
This ansatz has a time-dependent Tsallis $q$ maximum entropy ($q$-maxent) form, with the
time dependence represented in the parameters $\eta$, $\alpha$, $\delta$ and $\gamma$.
We then define
\be
  \label{eq:arg_Fq}
  \varphi =  1 - (1-q)(\alpha x^2 + \delta xy + \gamma y^2)\, ,
\ee
calculate the terms of the nonlinear Fokker-Planck equation (\ref{eq:NLFPE}) and obtain
the following expressions
\begin{subequations}
\label{eq:terms_NLFP}
\begin{align}
  \frac{\partial F}{\partial t} &= \dot{\eta} \varphi^\frac{1}{(1-q)} - \eta (\dot{\alpha} x^2 + \dot{\delta} xy + \dot{\gamma}y^2) \varphi^\frac{q}{(1-q)} \, ,
       \label{eq:dfdt} \\
 \frac{\partial^2 F^{2-q}}{\partial x^2} &= (2-q) \eta^{2-q} \left( -2\alpha \varphi^\frac{1}{(1-q)} + (2 \alpha x + \delta y)^2 \varphi^\frac{q}{(1-q)} \right) \, ,
        \label{eq:d2fdx2} \\
 \frac{\partial^2 F^{2-q}}{\partial y^2} &= (2-q) \eta^{2-q} \left( -2\gamma \varphi^\frac{1}{(1-q)} + (2 \gamma y + \delta x)^2 \varphi^\frac{q}{(1-q)} \right) \, ,
        \label{eq:d2fdy2} \\
 \pderiv{[(2ax + by)F]}{x} &= \eta \left[ 2a \varphi^\frac{1}{(1-q)} - (2ax + by) (2 \alpha x + \delta y) \varphi^\frac{q}{(1-q)} \right] \, , \label{eq:deriv2ax+by}\\
 \pderiv{[(2ay - bx)F]}{y} &= \eta \left[ 2a \varphi^\frac{1}{(1-q)} - (2ay - bx) (2 \gamma y + \delta x) \varphi^\frac{q}{(1-q)} \right] \, . \label{eq:deriv2ay-bx}
\end{align}
\end{subequations}
Next we substitute the right-hand side of the above equations (\ref{eq:terms_NLFP}) into the
NLFP equation (\ref{eq:NLFPE_QuadPot}) and, with some algebra, obtain the following set of
ordinary diferential equations for the time evolution of the
parameters $\eta$, $\alpha$, $\delta$ and $\gamma$
\begin{subequations}
\label{eq:ODE_parameters}
\begin{align}
  \frac{d \eta}{d t} &= 4 \eta a - 2 (2-q) D \eta^{2-q} (\alpha + \gamma) \, , \label{eq:detadt} \\
  \frac{d \alpha}{d t} &= -(2-q) D \eta^{1-q} \left( 4 \alpha^2 + \delta^2 \right) + 4 a \alpha - b \delta \, , \label{eq:dalphadt} \\
  \frac{d \gamma}{d t} &= -(2-q) D \eta^{1-q} \left( 4 \gamma^2 + \delta^2 \right) + 4 a \gamma + b \delta \, , \label{eq:dgammadt} \\
  \frac{d \delta}{d t} &= -4 (2-q) D \eta^{1-q} \delta (\alpha + \gamma) + 4 a \delta + 2 b (\alpha - \gamma) \, . \label{eq:ddeltadt}
\end{align}
\end{subequations}
Therefore the q-maxent ansatz (\ref{eq:TsallisAnsatz}) will be a solution of the NLFP equation (\ref{eq:NLFPE_QuadPot}),
provided that the functions $\eta (t)$, $\alpha (t)$, $\delta (t)$ and $\gamma (t)$ satisfy the
set of four coupled ordinary differential equations (\ref{eq:ODE_parameters}).

When we interpret the function $F_q(x_1, \cdots, x_N, t)$ (\ref{eq:TsallisF}) as a probability density
in phase space, or as a physical density of particles or other entities, we should require that
the norm $I$ of $F_q$ is finite, so that
\be
\label{eq:NormFq}
   I = \int F_q \: d x_1 d x_2 \cdots d x_N \le \infty\, .
\ee
For the density function (\ref{eq:TsallisAnsatz}) to have a finite norm,
in that expression, we should have $\alpha x^2 + \delta xy + \gamma y^2 = {\rm const. > 0}$,
determining the isodensity curves which should correspond to ellipses.
Therefore the quadratic form $\alpha x^2 + \delta xy + \gamma y^2$
has to be definite positive. Consequently, the discriminant
\be
    \varsigma = \alpha \gamma - \frac{\delta^2}{4}
\ee
has to be positive. It follows from (\ref{eq:ODE_parameters})
that the time derivative of the discriminant
\ben
   \label{eq:dvarsigmadt_2}
   \frac{d \varsigma}{d t} &=&
   [\delta^2 - 4 \alpha \gamma][ (2-q) D \eta^{1-q} (\alpha + \gamma) - 2 a ]\cr
   &=& 4 \varsigma [2a - (2-q) D \eta^{1-q} (\alpha + \gamma)]\, .
\een
We see that the value of the discriminant is not constant in time.
However, equation (\ref{eq:dvarsigmadt_2}) implies that the positive
character of $\varsigma$ is preserved under the time evolution
of the system. For the proposed $q$-statistical ansatz (\ref{eq:TsallisAnsatz}),
we find after some algebra that, for $q<1$, the norm (equation \ref{eq:NormFq}) is
\be
  \label{eq:normFq1_}
    I = \frac{\pi \eta}{ (2-q)\sqrt{ \alpha\gamma - \frac{\delta^2}{4} } }\, .
\ee
After some more calculation, it is also possible to verify using the
equations of motion (\ref{eq:ODE_parameters}) that
\be
 \label{eq:Const_I}
   \frac{dI}{dt} = \frac{\partial I}{\partial \eta}\frac{d\eta}{dt} +
      \frac{\partial I}{\partial \alpha}\frac{d\alpha}{dt} +
      \frac{\partial I}{\partial \gamma}\frac{d\gamma}{dt} +
      \frac{\partial I}{\partial \delta}\frac{d\delta}{dt} = 0\, ,
\ee
so that $I$ is a conserved quantity during the time evolution
of the system, as is to be expected.

A density $F({\bm x},t)$ governed by the partial differential equation
 (\ref{eq:NLFPE_QuadPot}) can be interpreted as describing
the distribution of a set of particles interacting via short range interactions,
performing overdamped motion under the drag effects due to a uniformly rotating medium,
and confined by an external harmonic potential. To see this, let us consider
the equation of motion of one individual test particle of this system
\be \label{particlemotion}
  m \ddot{\bm r} = - {\bm \nabla} W_{\rm int} - {\bm \nabla} W_{\rm ext}
  - \Gamma (\dot{\bm r} - \dot{\bm r}_R)\, ,
\ee
where $m$ is the mass of the test particle, $ W_{\rm int}$ is the
potential function associated with the forces acting on the
test particle due to the other particles of the system, $W_{\rm ext}$
is the external confining potential, and $\Gamma $ is a drag coefficient
describing the drag forces due to a resisting medium that rotates
uniformly with an angular velocity $\Omega$. Notice that the equation
of motion (\ref{particlemotion}) is expressed with respect to
an inertial reference frame (with cartesian coordinates $(x,y)$) and not
with respect to the rotating frame where the resisting medium is at rest.
With respect to the inertial frame, the local velocity $\dot{\bm r}_R$
of the medium has components $(- \Omega y, + \Omega x)$.

Since the interactions between the particles are short-range, we assume that
the potential function $W_{\rm int}$ is a function of the local
density $F$, that is $W_{\rm int} = {\cal D}(F)$. In the regime
of overdamped motion, equation (\ref{particlemotion}) becomes
\be \label{overdamped}
  \dot{\bm r} = - \frac{1}{\Gamma} \, {\bm \nabla} W_{\rm int} -
  \frac{1}{\Gamma} \, {\bm \nabla} W_{\rm ext} + \dot{\bm r}_R,
\ee
implying that the velocity $\dot{\bm r}$ of a particle in the system,
 at a given time, is completely determined by its location ${\bm r}$.
It can then be verified, after some calculations, that the continuity equation in configuration space,
$\partial F/ \partial t = - {\bm \nabla} (\dot {\bm r} F)$, describing
the evolution of the space density $F$ of a set of articles moving according to
the equation of motion (\ref{overdamped}),  is precisely
the NLFP equation (\ref{eq:NLFPE_QuadPot}), after the identifications
$\frac{{\cal D}}{\Gamma} = \frac{2-q}{1-q} D F^{1-q} $,
$\frac{W_{\rm ext}}{\Gamma} = a {\bm r}^2$, and $b= \Omega$.
\begin{figure}
  \includegraphics[width=95mm]{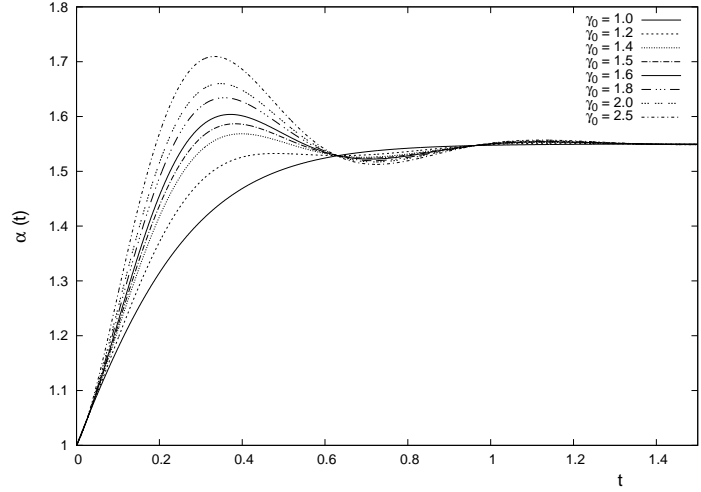}
  \caption{Evolution of the parameter $\alpha$ appearing in the
           time-dependent solution (\ref{eq:TsallisAnsatz})
           of the NLFP equation (\ref{eq:NLFPE_QuadPot}),
           for $q = 0.5$. The units employed are defined in terms
           of the constants $D$ and $b$ appearing in the NLFP equation.
           The parameter $\alpha$ has dimensions of inverse squared length and is
           measured in units of $\frac{b}{8D}$. The time $t$
           is measured in units of $\frac{4}{b}$.}
  \label{Fig:Alpha}
\end{figure}

\begin{figure}
  \includegraphics[width=95mm]{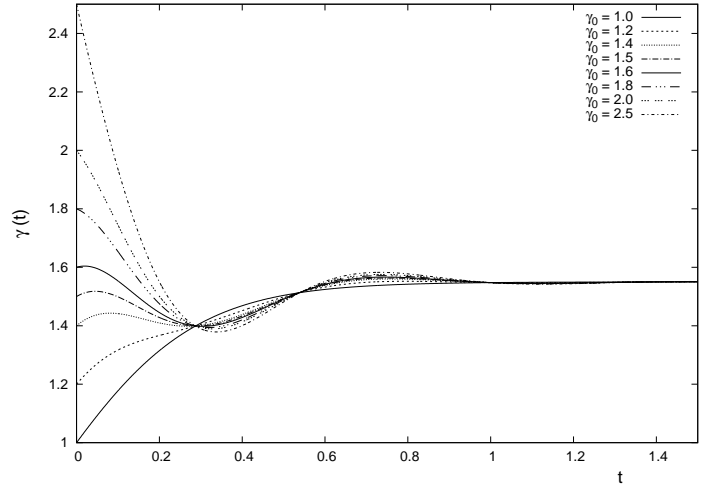}
  \caption{Evolution of the parameter $\gamma$
      appearing in the time-dependent solution (\ref{eq:TsallisAnsatz})
      of the NLFP equation, for $q = 0.5$. The parameter $\gamma$
           has dimensions of inverse squared length. The units employed are the same
           as in Figure 1.}
  \label{Fig:Gamma}
\end{figure}

\begin{figure}
  \includegraphics[width=95mm]{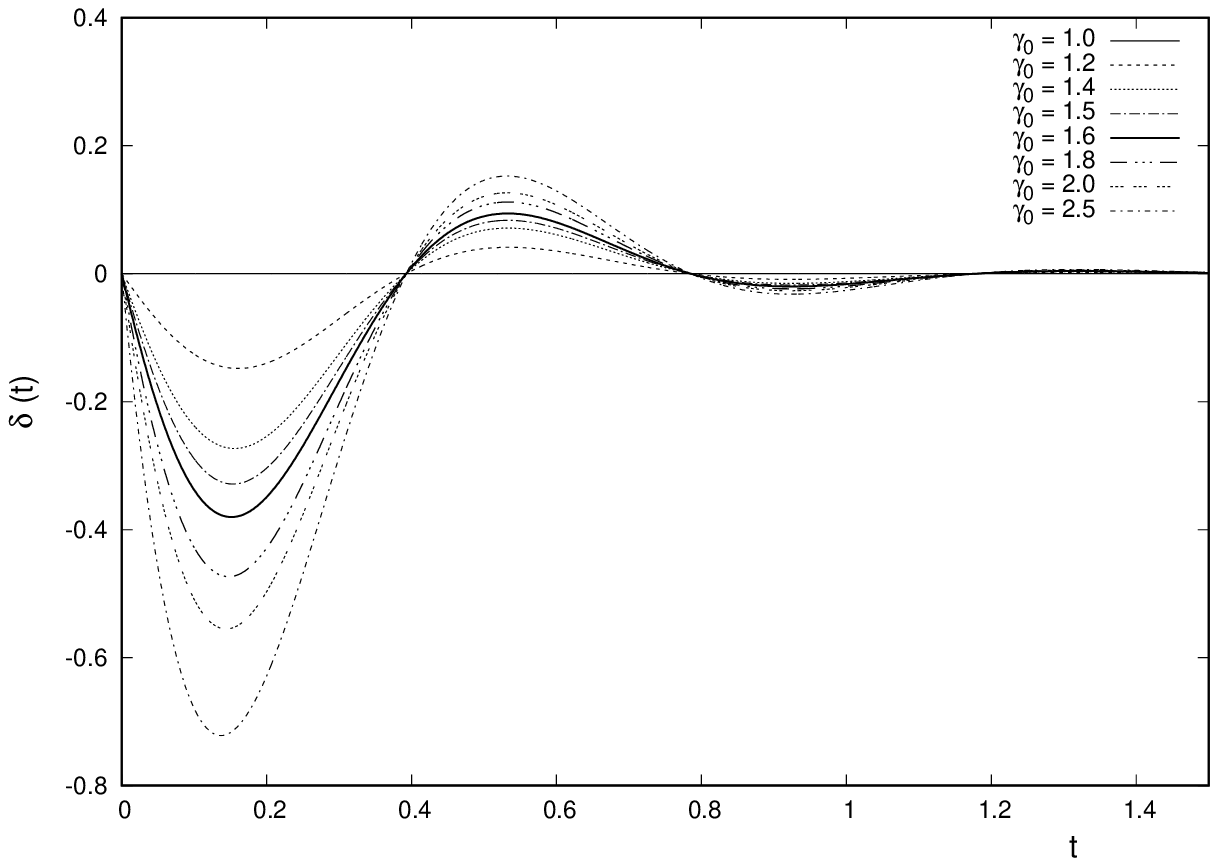}
  \caption{Evolution of the parameter $\delta$
      appearing in the time-dependent solution (\ref{eq:TsallisAnsatz})
      of the NLFP equation, for $q = 0.5$. The parameter $\delta$
           has dimensions of inverse squared length. The units employed are the same
           as in Figure 1.}
  \label{Fig:Delta}
\end{figure}

\begin{figure}
  \includegraphics[width=95mm]{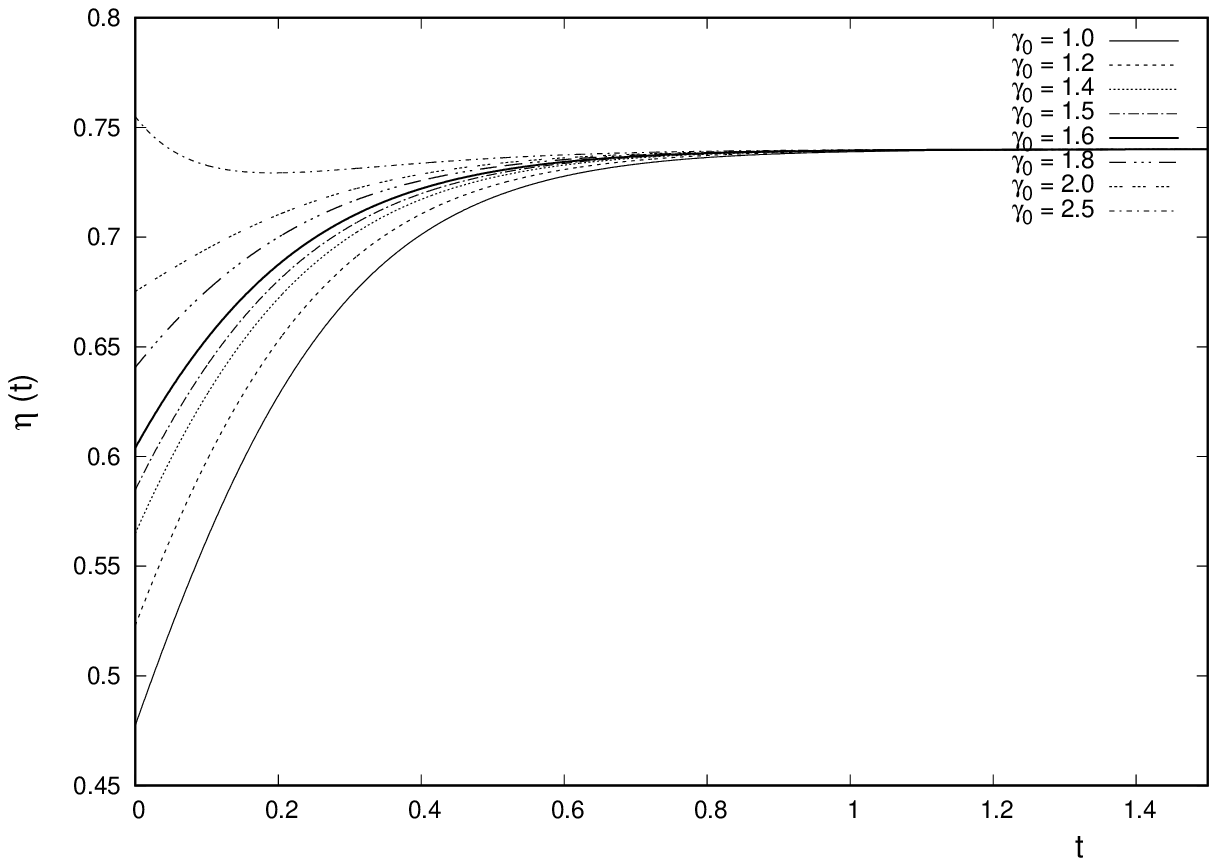}
  \caption{Evolution of the parameter $\eta$
      appearing in the time-dependent solution (\ref{eq:TsallisAnsatz})
      of the NLFP equation, for $q = 0.5$. The parameter $\eta$ is dimensionless.
      The time $t$ is measured in the same units as in Figure 1.}
  \label{Fig:Eta}
\end{figure}

An illustrative example of the time evolution of the $q$-Gaussian
solution (\ref{eq:TsallisAnsatz}) is provided in Figures~\ref{Fig:Alpha}-\ref{Fig:Eta}.
In these Figures, the
parameters $\alpha $, $\gamma$, $\delta$, and $\eta$, determining the
evolving size and shape of the two-dimensional $q$-Gaussian (\ref{eq:TsallisAnsatz}), are
depicted as a function of time. The different curves shown in
each Figure correspond to the NLFP equation (\ref{eq:NLFPE_QuadPot}), with $q= 0.5$, $D=0.5$,
$a= 1$, $b=4$, and different initial conditions.
The curves were therefore obtained from the numerical integration of the set of coupled
ordinary differential equations (\ref{eq:ODE_parameters}).
All solutions exhibited correspond to evolving densities normalized
to unity (that is, $I=1$. See equation (\ref{eq:normFq1_})). The initial conditions
are $\alpha_0 = 1$, $\delta_0 = 0$, with different initial values
of the parameter $\gamma$, as indicated in the Figures. The initial value of $\eta$ is calculated
from the initial values of the other three parameters, using the normalization
condition $I=1$.

It can be appreciated from Figures~\ref{Fig:Alpha}-\ref{Fig:Eta} that the different initial
densities considered (all having the same norm $I=1$) relax to the same
final stationary distribution (characterized by the same value of the norm).
This stationary distribution is rotationally symmetric. Consequently,
the initial asymmetry of the density tends to decrease as the evolution
takes place (the two axis of the isodensity curves tend to become equal
to each other). The oscillatory behavior of the parameter $\delta$, which takes
alternating signs as time advances, indicates that the asymmetric density
rotates as the evolution proceeds. Note that at the times when $\delta =0$
the axis of the isodensity curves are parallel to the coordinate axis. This
happens at approximately regular time intervals, indicating that the elliptical
isodensity curves rotate at an approximately constant mean angular velocity.
The oscillatory behavior associated with the rotation affects the other
variables (besides $\delta$) as well, which also exhibit oscillations whose
amplitudes tend to decrease as the density function $F$ relaxes towards the
stationary one.


\section{Conclusions}

We  investigated the main properties of multi-dimensional NLFP
equations involving curl drift forces.
We considered drift force fields comprising both
an irrotational term ${\bm G}$ derived from a potential function $V({\bm x})$
and a curl, non-gradient term ${\tilde {\bm K}}$. We determined the requirements
that the two parts ${\bm G}$ and ${\tilde {\bm K}}$ of the drift
field have to satisfy, in order for the corresponding NLFP equation
to admit a stationary solution of the $q$-maxent form (that is, a
$q$-exponential of the potential function $V({\bm x})$ associated with the
gradient component of the drift force). We found that this kind of stationary
solution exists for a continuous range of values of the parameter $\beta$
if and only if, the curl part ${\tilde {\bm K}}$ is divergenceless
and the curl part is orthogonal to the gradient part ${\bm G}$. We also proved that NLFP
equations admitting a stationary solution also verify an $H$-theorem,
in terms of an appropriate linear combination of the $S_q$ entropic functional
and the mean value of the potential $V$. Finally, we studied exact analytical
time-dependent solutions of a two dimensional NLFP equation, describing a system
of interacting particles in an overdamped motion regime, under the drag effects
originating on a uniformly rotating medium. The connection between rotation and
NLFP equations with curl forces, combined with the connection between
$q$-thermostatistics and self-gravitating systems, indicates that those evolution equations may
have applications in geophysical and astrophysical problems. Previous successful
physical applications of NLFP equations also suggest that experimental implementations
involving rotating granular materials may also be worth exploring.

Another potential field of application of the NLFP dynamics, investigated
in the present work, is the space-time behavior of some biological systems \cite{C2008}.
Diffusion processes are useful to model the spread of biological populations \cite{M2004,F2007}.
Nonlinear diffusion equations have been proposed, as effective descriptions
of the interaction between the members of a diffusing biological population
\cite{MRMLN2015,NS81,TFCP2007}.
On the other hand, drift terms can be used to describe other effects
on the motion of the individuals. In this biological context, since
the ``forces" are not fundamental but rather the effective result
of a set of complex circumstances, it is to be expected that non-gradient
forces can be relevant. NLFP equations with non-gradient drift fields may also be
useful in connection with the generalized Boltzmann machine approach (based
on a $q$-generalization of simulated annealing \cite{TS96}) to neural network
models of memory \cite{WDC2009}, when considering asymmetric neural interactions.
Any further developments along these or related lines will be very welcome.

\vskip 2\baselineskip

{\large\bf Acknowledgments}

\vskip \baselineskip
Two of us (R.W. and A.R.P.) acknowledge warm hospitality and access to computational
facilities at the Centro Brasileiro de Pesquisas Fisicas and the National Institute
of Science and Technology of Complex Systems,  partially supported by the Brazilian
Agencies CNPq, FAPERJ and CAPES. One of us (C.T.) also acknowledges partial financial
support from the John Templeton Foundation (USA).


\end{document}